\documentclass{ws-mpla}

\newif\ifgeneratebbl
\generatebbltrue

\usepackage{amsmath,amssymb}
\renewcommand{\thesection}{\arabic{section}}
\renewcommand{\theequation}{\thesection.\arabic{equation}}
\newcommand{\eq}[1]{Eq.~(\ref{#1})}
\newcommand{\fig}[1]{Fig.~{\ref{#1}}}
\newcommand{\be}{\begin{equation}}
\newcommand{\ee}{\end{equation}}
\newcommand{\bea}{\begin{eqnarray}}
\newcommand{\eea}{\end{eqnarray}}
\newcommand{\ST}{Slavnov--Taylor }
\newcommand{\YM}{Yang--Mills }
\newcommand{\DS}{Dyson--Schwinger }
\newcommand{\BS}{Bethe--Salpeter }
\newcommand{\w}{\omega}
\newcommand{\e}{\varepsilon}
\newcommand{\al}{\alpha}

\newcommand{\ga}{\gamma}
\newcommand{\G}{\Gamma}
\newcommand{\de}{\delta}

\newcommand{\si}{\sigma}

\newcommand{\La}{\Lambda}
\newcommand{\ka}{\kappa}

\newcommand{\ha}{\frac{1}{2}}
\newcommand{\pd}{\partial}

\newcommand{\cd}{{\cal D}}
\newcommand{\cs}{{\cal S}}

\newcommand{\s}[2]{{#1}\!\cdot\!{#2}}
\newcommand{\ov}[1]{\overline{#1}}

\newcommand{\im}{\mathrm{i}}

\newcommand{\prb}{Phys.\ Rev.\ B}
\newcommand{\prd}{Phys.\ Rev.\ D}
\newcommand{\nat}{Nature\ Physics}

\usepackage{upquote}

\begin{document}

\markboth{Carina Popovici}
{Dyson--Schwinger approach to strongly coupled theories}

%%%%%%%%%%%%%%%%%%%%% Publisher's Area please ignore %%%%%%%%%%%%%%
\catchline{}{}{}{}{}
%%%%%%%%%%%%%%%%%%%%%%%%%%%%%%%%%%%%%%%%%%%%%%%%%%%%%%%%%%%%%%%%%%%

\title{\bf DYSON--SCHWINGER APPROACH TO STRONGLY COUPLED THEORIES}

\author{\footnotesize CARINA POPOVICI}
\address{
Institut f\"ur Theoretische Physik, Justus-Liebig-Universit\"at
Giessen, Heinrich-Buff-Ring 16, 35392 Giessen, Germany \\
carina.popovici@theo.physik.uni-giessen.de}

\maketitle

\pub{Received (Day Month Year)}{Revised (Day Month Year)}

\begin{abstract}

Although nonperturbative functional methods are often associated with
low energy Quantum Chromodynamics, contemporary studies indicate that
they provide reliable tools to characterize a much wider spectrum of
strongly interacting many-body systems. In this review, we aim to
provide a modest overview on a few notable applications of \DS
equations to QCD and condensed matter physics. After a short
introduction, we lay out some formal considerations and proceed by
addressing the confinement problem. We discuss in some detail the
heavy quark limit of Coulomb gauge QCD, in particular the simple
connection between the nonperturbative Green's functions of \YM theory
and the confinement potential. Landau gauge results on the infrared
\YM propagators are also briefly reviewed. We then focus on less
common applications, in graphene and high-temperature
superconductivity. We discuss recent developments, and present
theoretical predictions that are supported by experimental findings.

\keywords{Dyson-Schwinger equations; confinement;
  graphene; high-temperature superconductivity}
\end{abstract}

\ccode{PACS Nos.: 71.10.-w,12.38.Aw,72.80.Vp}

%71.10.-w Theories and models of many-electron systems
%12.38.Aw Quark confinement
%72.80.Vp Electronic transport in graphene

\section{Introduction}	

Quite generally, one of the challenges of modern theoretical physics
is to explain the dynamics of quantum systems. Among them,
fermion-fermion interactions are very specific and generate a large
number of effects --- from quark confinement in hadron physics, to
quark gluon plasma in heavy ion collisions, from superconductivity in
metals to band structure in crystals. Many different methods have been
devised to tackle the many-body problem, under a variety of
circumstances and at different energy scales.  In high energy QCD,
where asymptotic freedom guarantees that the coupling between quarks
and gluons can be treated as a small parameter, perturbation theory
has been successfully applied. At intermediate and low momenta,
different methods are required since the coupling increases as we go
to lower energies and perturbation theory alone cannot give a good
description of the theory. One possibility is to employ continuum
functional techniques such as renormalization
group\cite{Pawlowski:2005xe} and \DS
equations.\cite{Alkofer:2000wg,Fischer:2006ub} Alternatively,
numerical lattice simulations represent a potentially exact method,
however extrapolation to chiral and infinite volume limits is often a
numerical challenge.\cite{Greensite:2003bk,Colangelo:2010et} The
situation is very similar in condensed matter electronic systems,
where an equivalent program has been
implemented.\cite{anderson84,Shankar:1993pf,drula2009,drni2012}

This review is organized as follows. In section~\ref{sec:DS} we
briefly review the general features of \DS equations, and proceed in
section~\ref{sec:QCD} by discussing the confinement problem, with
particular emphasis on the heavy quark limit of Coulomb gauge QCD. A
brief overview of the progress made on infrared propagators of Landau
gauge \YM theory is also attempted. Phenomenological findings are not
discussed here, since they have been presented in many other works,
see for example Ref.~\refcite{Roberts:2012sv} for a recent review. In
section~\ref{sec:QED3} we turn to planar condensed matter systems. We
exemplify the \DS approach to strongly interacting many electron
systems with two case studies, graphene and high temperature
superconductors (HTSs). Our choice is motivated by the fact that these
systems share a common aspect, namely, they can be described by
effective quantum field theoretical models that inherit features of
both (2+1) dimensional Quantum Electrodynamics (QED$_3$) and its four
dimensional counterpart. We present results obtained from the \DS
equations, and where available the supporting experimental
evidences. At the more formal level, it is also worth mentioning the
similarities between the propagators of \YM theory and QED$_3$, i.e.,
the possibility of power law behavior in the deep infrared. A short
summary is presented in section~\ref{sec:concl}.

%%%%%%%%%%%%%%%%%%%%%%%%%%%%%%%%%%%%%%%%%%%%%%%%%%%%%%%
%%%%%%%%%%%%%%%%%%%%%%%%%%%%%%%%%%%%%%%%%%%%%%%%%%%%%%%

\section{\DS formalism}
\label{sec:DS}

\DS equations built up an infinite set of coupled non-linear integral
equations that relate the various Green's functions of a quantum field
theory.  From a practical point of view, defining a tractable problem
requires the ability to reduce these equations to a closed subset,
e.g. by making controlled ans\"atze for the higher order Green's
functions. In general the resulting coupled integral equations are
solved numerically, nevertheless it is possible that in particular
setups, such as heavy quark sector of Coulomb gauge QCD (under certain
truncations), or infrared limit of Landau gauge \YM theory, exact
analytic solutions become available. These aspects will be discussed
in more detail in the next section.

The derivation of the \DS equations rests on the idea that the
generating functional of a theory
\be
Z[J]=\int\cd\Phi\exp{\left\{\im\cs + \im\cs_{s}\right\}}
\label{eq:Z}
\ee
is invariant under the variation of a generic field $\Phi_{\al}$:
\be
\int\cd\Phi\frac{\delta}{\delta \im \Phi_{\al}}
\exp\left\{\im\cs+\im\cs_{s}\right\}
=0,
\label{eq:funcder}
\ee
where $\cs$ is the action of the problem at hand, $\cs_{s}$ the
corresponding source term and $\cd\Phi$ denotes the integration over
all fields. To ensure the validity of this equation, translational
invariance of the measure $\cd\Phi$ is assumed. From \eq{eq:funcder},
one can derive the \DS equation for any $n$-point Green's function, by
taking further functional derivatives with respect to the fields and
omitting those terms which vanish when the sources are set to zero.

\begin{figure}[t]
\centering\includegraphics[width=1.0\linewidth]{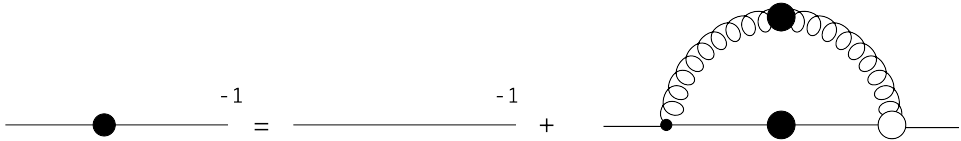}
\vspace*{8pt}
\caption{\protect\label{fig:gapeq} \DS equation for a generic fermion
propagator in a gauge theory. Filled circles denote dressed
propagators, the empty circle denotes the dressed fermion-gauge boson
vertex, the spring denotes the gauge propagator and solid lines stand
for the fermion propagator.}
\end{figure}

Among the most investigated \DS equations is the fermion gap equation,
which, regardless of the underlying gauge theory, has the
diagrammatical representation shown in \fig{fig:gapeq}. It essentially
states that the inverse of the nonperturbative fermion propagator
equals the sum between the inverse of the bare fermion propagator and
the fermion self energy, which in turn contains dressed fermion and
gauge boson propagators, and one bare and one dressed fermion-gauge
boson vertex. Formally, the fermion gap equation appears in systems as
discrepant as QCD, graphene and superconductors, whereby the raw form
\fig{fig:gapeq} has to be customized to the particular theory under
consideration.

In addition to formal analogies, there are also important qualitative
similarities between the systems that we are going to examine. An
example is the possibility of scaling laws for the propagators in the
deep infrared --- a feature shared by both Landau gauge \YM theory and
QED$_3$ (including high temperature superconductors, which can be
described with a model that has properties reminiscent of QED$_3$).
\cite{Fischer:2006ub,fial2004,Bonnet:2011hh}

%------------------------------------------------------------------
%------------------------------------------------------------------

%%%%%%%%%%%%%%%%%%%%%%%%%%%%%%%%%%%%%%%%%%%%%%%%%%%%%%%
%%%%%%%%%%%%%%%%%%%%%%%%%%%%%%%%%%%%%%%%%%%%%%%%%%%%%%%

\section{Confining solutions from the \DS equations}
\label{sec:QCD}

Confinement, i.e., the experimental evidence that quarks and gluons
must be confined into color singlet hadronic states, represents, along
with chiral symmetry breaking, one of the theoretical pillars of low
energy hadron physics. A rigorous understanding of this phenomenon,
i.e., identifying the specific mechanisms that act at the level of the
underlying (gauge dependent) Green's functions, is to date still an
open problem.

As is well known, in order to overcome the difficulties related to the
definition of the generating functional \eq{eq:Z}, the QCD action
needs to be supplemented by a gauge fixing term, along with the
associated ghost fields.  Among various options, Coulomb gauge $\pd_i
A_i^a=0$ has properties that qualify it as an efficient choice to
study nonperturbative phenomena: in this gauge, Gauss law is naturally
built in such that gauge invariance is fully accounted for, the total
color charge is conserved and
vanishing,\cite{Zwanziger:1998ez,Reinhardt:2008pr} and most
importantly, a natural picture of confinement emerges. Furthermore,
Coulomb gauge allows for different manifestations of confinement
depending on the specific formalism, as shall be discussed below.

In the Hamiltonian formalism, Coulomb gauge is implemented
simultaneously with Weyl gauge $A_0=0$\footnote{Note, however, that
imposing Weyl gauge leads to the fact that Gauss law is no longer
arising from the equations of motion, but has to be imposed as an
external
constraint.}.\cite{Feuchter:2004gb,Feuchter:2004mk,Reinhardt:2004mm,Szczepaniak:2001rg,Szczepaniak:2005xi}
In this approach, the \DS equations for the ghost and spatial gluon
propagators have been solved, analytically in the infrared and
numerically in the whole momentum regime
\cite{Feuchter:2004mk,Reinhardt:2004mm,Schleifenbaum:2006bq,Epple:2006hv,Epple:2007ut}
(in Ref.~\refcite{Reinhardt:2011hq} recent results at finite
temperature are presented). Similar to Landau gauge scaling
solutions,\cite{vonSmekal:1997is,vonSmekal:1997vx,Lerche:2002ep} the
ghost and gluon dressing functions exhibit power law behaviors,
$(p^2)^{\de_{gh}}$ and $(p^2)^{\de_{gl}}$. A general infrared power
law analysis has yielded the following relation between the infrared
exponents (with Landau and Coulomb gauge differing only in the number
of dimensions $d$) \cite{Zwanziger:2001kw,Schleifenbaum:2006bq}
\be
\de_{gl}+2\de_{gh}=\frac{4-d}{2}.\label{eq:scagen}
\ee
In the case $d=3$, which corresponds to equal-time Green's functions
in Coulomb gauge, two power law exponents $\kappa:=-\de_{gh}$ have
been found, $\ka^{(1)}=0.398$ and $\ka^{(2)}=1/2$
\cite{Schleifenbaum:2006bq}. The energetically favored solution is the
most singular, which produces a ghost propagator dressing function
diverging as $1/|\vec k|$. In turn, this gives rise to a strictly
linearly rising static quark potential at large
distances.\cite{Epple:2006hv} The case $d=4$ trivially leads to the
Landau gauge scaling relation $\de_{gh}=-\de_{gl}/2$
\cite{Zwanziger:2001kw,Lerche:2002ep} (see also the discussion at the
end of this section).

In the functional formalism (as applied to Coulomb gauge QCD), apart
from the ghost and transversal spatial gluon propagator, there is a
third propagator, the temporal gluon propagator, which plays an
important role in uncovering the confining potential between (static)
quarks. To see how this comes about, let us first briefly review the
main characteristics of the functional formalism.

In general, functional \DS studies in Coulomb gauge are plagued by the
so-called energy divergence problem, i.e., the unregulated divergences
generated by ghost loops, as in the one-loop integral $\int dk_{0}\int
d^3\vec k [(\vec k-\vec p)^2 \vec k^2]^{-1}$. The usual dimensional
regularization fails in this case, although the full set of such loops
should cancel in the \DS equations. While such cancellations have been
isolated up to two loops in perturbation
theory,\cite{Andrasi:2005xu,Watson:2007mz,Leibbrandt:1996tn,Heinrich:1999ak}
they are exceedingly difficult to pin down in the full tower of \DS
equations.  This problem can be bypassed by converting to first order
functional formalism, i.e., linearizing the chromoelectric term in the
action via the auxiliary field $\vec\pi$ \cite{Zwanziger:1998ez}
\be
\exp{\left\{\im\int d^4x\ha\s{\vec{E}}{\vec{E}}\right\}}
=\int\cd\vec{\pi}\exp{\left\{\im\int d^4x\left[-\ha
\s{\vec{\pi}}{\vec{\pi}}-\s{\vec{\pi}}{\vec{E}}\right]
\right\}}.
\label{eq:piintro}
\ee
The $\vec{\pi}$ field is subsequently split into transverse and
longitudinal components (details are provided in
Ref.~\refcite{Watson:2006yq}). With this new fields, the QCD action is
rewritten such that the ghost field, i.e., the Faddeev-Popov
determinant, cancels against the inverse functional determinant that
stems from resolving the chromodynamical equivalent of the Gauss law.
The resulting action contains only the 'would-be-physical' degrees of
freedom, the transverse $\vec{A}$ and $\vec{\pi}$ fields (which
classically would correspond to the configuration variables and their
momentum conjugates), whereas the unphysical ghosts are formally
eliminated. The shortcoming, however, is that we end up with a
nonlocal Lagrangian that is unsuited for practical
applications.\cite{Zwanziger:1998ez,Watson:2006yq} Nevertheless, the
nonlocal formulation can provide important guidelines to the local
theory.

It is important to realize that identifying the mechanisms that lead
to the cancellations of the unphysical components is essential for a
thorough understanding of the theory. For example, the (unphysical)
energy-independent ghost loop mentioned above should be canceled by
the temporal component of the gluon propagator, and this implies that
the temporal gluon itself must have an energy-independent
part.\cite{Cucchieri:2000hv} This argument is supported by lattice
results \cite{Quandt:2008zj} and analytic findings
\cite{Watson:2010cn}. In the limit of the heavy quark mass, this
information has been used as input to derive a relation between the
temporal gluon propagator and the nonperturbative scale associated
with confinement (the string tension).\cite{Popovici:2010mb} We will
return to this relation below.

%------------------------------------------------------------------
\begin{figure}[t]
\centering\includegraphics[width=0.69\linewidth]{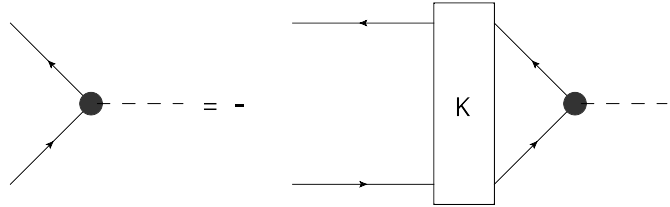}
\vspace*{8pt}
\caption{\protect\label{fig:bseq} Homogeneous \BS equation for
quark-antiquark bound states. Solid lines represent the (dressed)
quark propagators, dashed lines denote the corresponding bound state,
and blobs denote the \BS vertex function. The box represents the \BS
kernel which in the heavy quark limit reduces to the exchange of a
single temporal gluon.}
\end{figure}
%------------------------------------------------------------------
Explicitly, heavy quark limit refers in this case to a heavy quark
mass expansion of the QCD action at leading order,\footnote{The QCD
action is expanded in powers of the inverse quark mass by means of a
heavy quark transformation adapted from
Ref.~\refcite{Neubert:1993mb}.} with the truncation of the \YM sector
to include only dressed propagators.\footnote{A similar Coulomb gauge
truncation scheme with arbitrary quark mass has also been studied, and
which explicitly reproduces the heavy quark limit presented
here.\cite{Watson:2011kv,Watson:2012ht}} In this framework, a full
nonperturbative study of the gap and \BS equations has been
performed. Solving the gap equation\footnote{The full gap equation
within Coulomb gauge first order formalism (without the mass
expansion) has been derived in Ref.~\refcite{Popovici:2008ty}.}
(generically represented in \fig{fig:gapeq}) yields the following
solution for the heavy quark propagator:
\bea
W_{\ov{q}q}(k_0)=\frac{-\im}{\left[k_0-m-
{\cal I}_r+\im\e\right]},
\label{eq:quarkpropnonpert}
\eea
where $m$ denotes the quark mass and ${\cal I}_r$ is an (implicitly
regularized) constant. Since this propagator has a single pole in the
complex $k_0$ plane, it follows that the closed quark loops are
suppressed, and this implies that the theory is quenched in the heavy
mass limit:
\be
\int dk_0 \frac{1}{\left[k_0-m-
{\cal I}_r+\im\e\right]\left[k_0+p_0-m-
{\cal I}_r+\im\e\right]}=0.
\ee
By using the associated \ST identity, it is easy to show that the
temporal quark-gluon vertex remains nonperturbatively bare, $\G_{\bar
qq A_0} \to \G_{\bar qqA_0}^{(0)}= gT^a$, with $T^a$ being the
hermitian generators of the gauge group. Moreover, the kernel of the
\BS equation, depicted in \fig{fig:bseq}, reduces to a single gluon
exchange:
\be
K(k)=\G^{a}_{\bar qqA_0}\, W_{00}^{ab}(\vec k)\,\G^{Tb}_{q \bar qA_0},
\label{eq:kerqq}
\ee
where $W_{00}(\vec k)$ is the temporal gluon propagator (a complete
description and derivation is provided in
Ref.~\refcite{Popovici:2010mb}). With the kernel \eq{eq:kerqq}, the
bound state energy between a quark and an antiquark at leading order
in the mass expansion reads:
\be
P_{\bar qq}=g^2\int_{r}\frac{d \vec{\w}} {(2\pi)^3} 
W_{00}(\vec{\w})
\left[C_F-C_Me^{\im\vec{\w}\cdot\vec{x}}\right],
\label{eq:p0sol}
\ee
where the index '$r$' denotes an implicitly regularized integral,
$C_F$ is the Casimir factor, $C_M$ denotes an (unknown) color factor
assigned to the quark-meson vertex, and $|\vec x|$ represents the
separation between the quark and the antiquark. To identify $C_M$, we
recall that since a quark cannot live as asymptotic
state,\cite{Reinhardt:2008pr} the bound state energy $P_{\bar qq}$
must be linearly rising, or otherwise the energy is infinite and the
quark-antiquark system is not allowed. With a temporal gluon dressing
function more divergent than $1/|\vec\w|$, it follows that $C_M=C_F$
and hence a finite solution of the \BS equation exists only for color
singlet states. Assuming that in the infrared
\be
W_{00}(\vec\w)=\frac{X}{\vec{\w}^4},\label{eq:irtemp}
\ee
where $X$ is a combination of constants (and $g^2 X$ is a
renormalization group invariant \cite{Zwanziger:1998ez}),
\eq{eq:p0sol} reduces to
\be
P_{\bar qq}:=\si |\vec{x}|=\frac{g^2C_FX}{8\pi}|\vec{x}|.
\label{eq:enbarqq}
\ee
This result provides the link (at least at leading order) between the
physical string tension $\si$ and the nonperturbative \YM sector of
QCD. It corresponds to the only physical solution (a color singlet
bound state of a quark and an antiquark) or else the energy of the
system is divergent.  A similar calculation performed for the diquark
\BS equation has shown that diquarks are confined for $N_c=2$ colors,
which corresponds to a confined, antisymmetric bound state of two
quarks (the $SU(2)$ baryon), and otherwise there are no physical
states. Finally, a thorough investigation of the \DS equation for the
full nonperturbative four-point quark-antiquark Green's functions has
evidenced the separation of the physical and unphysical singularities
of the Green's function, and that the physical pole coincides with the
pole \eq{eq:p0sol} of the homogenous \BS
equation.\cite{Popovici:2011yz} An analogous investigation for the
diquark system has shown that the resonant pole of the four-point
Green's function is attached to the only physical solution, for
$N_c=2$ colors, corresponding to color antisymmetric and flavor
symmetric configuration.

%-------------------------------------------------------
\begin{figure}[t]
\centering\includegraphics[width=0.95\linewidth]{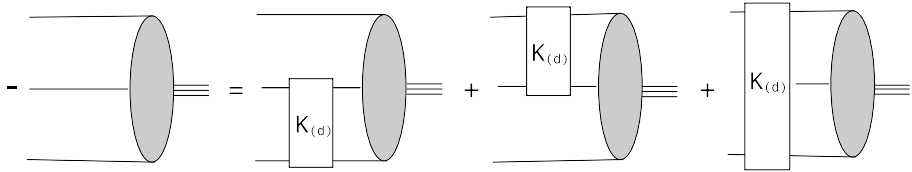}
\vspace*{8pt}
\caption{\protect\label{fig:fadeq} Faddeev equation for three-quark
bound states. The box depicts in this case the diquark kernel, the
ellipse denotes the Faddeev vertex function, and the triple lines
represent the three-quark bound state. Same conventions as in
\fig{fig:bseq} apply.}
\end{figure}
%------------------------------------------------------------------

Following the same approach, a similar analysis has been carried out
for three quark systems by means of Faddeev equation, depicted in
\fig{fig:fadeq}.\cite{Popovici:2010ph} Since in the quark-diquark
model the binding energy between quarks is assumed to be provided
essentially by two-quark correlations,\cite{Reinhardt:1989rw} in the
Faddeev equation only the permuted two-quark kernels $K_{(d)}$ are
employed, whereas the three-body irreducible diagrams are
neglected. Just like for $\bar qq$ states, the diquark kernel reduces
to the ladder approximation,
\be
K_{(d)}(k)= 
\G_{\bar qqA_0}^{a}\, W_{00}^{ab}(\vec k)\, \G_{\bar qqA_0}^{b}.
\label{eq:kernel}
\ee
By working in the symmetric case (equal spatial separations between
quarks), the following three-quark bound state energy has been
determined as a solution of the Faddeev equation
\be
P_{3q}=3m+\frac{3}{2}g^2\int_{r}\frac{d \vec{\w}} {(2\pi)^3} 
W_{00}(\vec{\w})
\left[C_F-2C_Be^{\im\vec{\w}\cdot\vec{x}}\right],
\label{eq:p1sol}
\ee
where $C_B$ denotes the color factor attached to the quark-baryon
vertex. As before, the only possibilities are either that the $3q$
system is confined, or it is physically not allowed. An infrared
confining solution requires that the condition $C_B=C_F/2$ is
satisfied, provided that the temporal gluon propagator is infrared
enhanced. This implies that the baryon is a color singlet bound state
of three quarks and otherwise the energy of the system is
infinite. Inserting $C_B=C_F/2$, along with the infrared temporal
gluon propagator \eq{eq:irtemp}, into \eq{eq:p1sol}, yields the
following result for the bound state energy of the three quark system:
\be
P_{3q}=3m+\frac{3}{2}\frac{g^2C_FX}{8\pi}|\vec{x}|.
\label{eq:enqqq}
\ee
As in the case of $\bar qq$ systems, the 'string tension', i.e., the
coefficient of the three-body linear confinement term, is directly
related to the nonperturbative temporal gluon propagator. Comparing
the above result with \eq{eq:enbarqq}, it appears that the string
tension corresponding to the $qqq$ system is 3/2 times that of the
$\bar qq$ system. The $3m$ term is a reminiscence of the heavy quark
transformation which enters the Faddeev equation via the heavy quark
propagator, \eq{eq:quarkpropnonpert}.

Let us now turn our attention towards Landau gauge $\pd_{\mu}
A_{\mu}^a=0$.  Unlike Coulomb gauge, where one is faced with severe
technical difficulties introduced by noncovariance, Landau gauge
remains popular due to its covariance, but also because of certain
technical simplifications (the ghost-gluon vertex remains
approximately bare in the infrared \cite{Schleifenbaum:2004id}). In
this gauge, the coupled system of \DS equations for ghost and gluon
propagators has been solved,\cite{Lerche:2002ep,Fischer:2002hna} see
also the discussion above \eq{eq:scagen}. It was found that the gluon
propagator vanishes at zero momentum, whereas the ghost propagator is
infrared enhanced, in agreement with the ghost dominance picture of
Gribov and
Zwanziger,\cite{Gribov:1977wm,Zwanziger:1995cv,Zwanziger:1998ez} and
Kugo and Ojima.\cite{Kugo:1979gm,Kugo:1995km,Watson:2001yv} With the
truncation to employ a bare ghost-gluon vertex, the scaling
coefficient $\ka \approx 0.595$ has been calculated independently in
Refs.~\refcite{Zwanziger:2001kw} and
\refcite{Lerche:2002ep}.\footnote{In fact, it turns out that this
result changes very little if the vertex is dressed in the
infrared.\cite{Lerche:2002ep}} By combining informations stemming from
both \DS and renormalization group equations, it has been shown that
the Landau gauge scaling solution is unique (however, this should not
be confused with the uniqueness of the solution
itself).\cite{Fischer:2009tn}

On the lattice, it eventually become clear that the gluon propagator
is not vanishing at small momentum; instead, a massive behavior has
been seen, although some papers were still in agreement with a
vanishing gluon propagator, see for example
Ref.~\refcite{Vandersickel:2012tz} and references therein.  These
findings support the so-called decoupling solution --- a second
solution derived with \DS methods, where the gluon and ghost propagate
quite differently: at small momenta, the gluon propagator becomes
finite instead of going to zero, whereas the ghost dressing remains
finite.\cite{Aguilar:2008xm,Dudal:2007cw,Dudal:2008sp} In order to
understand the connection between the two solutions, it is necessary
to examine the renormalization of the ghost propagator. A finite ghost
dressing function at zero momentum generates a continuous set of
decoupling solutions, whereas if an infinite ghost dressing at zero
momentum is employed, the scaling solution is
recovered.\cite{Fischer:2008uz} Finally, we note that a numerical
analysis of gluon and ghost propagators in the complex momentum plane
has been recently carried out.\cite{Strauss:2012dg}

Importantly, both the decoupling and scaling solutions fulfill the
criterion for a confining Polyakov loop potential, meaning that quarks
are confined.\cite{Braun:2007bx} This criterion depends only on the
asymptotic part, and as it turns out, in the actual calculations the
dynamics of the system is driven by the non-perturbative mid-momentum
regime (where both solutions agree), and not the deep infrared.

%%%%%%%%%%%%%%%%%%%%%%%%%%%%%%%%%%%%%%%%%%%%%%%%%%%%%%
%%%%%%%%%%%%%%%%%%%%%%%%%%%%%%%%%%%%%%%%%%%%%%%%%%%%%%

\section{Manifestations of (2+1) dimensional QED in condensed matter
physics}
\label{sec:QED3}

Although when studying gauge theories it seems natural to consider a
theory set in $3+1$ dimensions, it is nevertheless possible that some
insight might be gained by restricting to $2+1$ dimensions. An example
is QED$_3$, which apart from having intrinsically interesting
features, such as superrenormalizability,\cite{Roberts:1994dr} is also
attractive for its ability to describe the behavior of certain
materials that lay at the border between particle and condensed matter
physics, such as graphene and high temperature superconductors. The
quantum field theoretical modeling of these materials has provided a
description of phenomena that are nonperturbative in essence,
e.g. dynamical mass generation via chiral symmetry breaking.

Before turning our attention to condensed matter systems, we note that
the infrared behavior of Landau gauge QED$_3$ propagators is given by
power laws, just like the scaling solutions of Landau gauge \YM theory
(see also the discussion from the previous section).  The QED$_3$
power laws have been derived from an infrared analysis of the coupled
system of fermion and photon \DS equations.\cite{fial2004}

%%%%%%%%%%%%%%%%%%%%%%%%%%%%%%%%%%%%%%%%%%%%%%%%%%%%%%%%%%%%%%
%%%%%%%%%%%%%%%%%%%%%%%%%%%%%%%%%%%%%%%%%%%%%%%%%%%%%%%%%%%%%%

\subsection{Graphene}
\label{subsec:graphene}

Graphene is a monatomic layer of carbon atoms arranged on a honeycomb
lattice.\cite{CastroNeto:2009zz,Peres2010,Sarma2011} Its remarkable
electronic properties, such as unconventional quantum Hall
effect,\cite{Gusynin2005,Zhang2005} Klein tunneling
\cite{Katsnelson2006,Cheianov2006} or charge
confinement,\cite{Rozhkov2011} qualify it as one of the most promising
materials for future nanoscale devices. Its synthesis in 2004
\cite{Novoselov2004,Novoselov2005} was awarded the Nobel Prize in
physics in 2010.

The fundamentally new characteristics of this system are theoretically
explicable in terms of the low energy behavior of the charge carriers:
instead of the familiar Schr\"odinger equation, the electrons in
graphene are described by the relativistic Dirac equation for massless
fermions.\cite{Wallace1947,Berger2006} The resulting single particle
dispersion relation is linear in the momentum, $ E_{k}=\pm v_F |\vec
k|$, where $v_F$ is the Fermi velocity (about $300$ times smaller than
the speed of light) and $\vec k = (k_x, k_y)$ is the fermion momentum,
measured relative to the inequivalent corners of the Brillouin zone.

From a technological point of view, graphene based nanoscale devices
require the ability to control the so-called charge confinement, i.e.,
the clustering of electronic charge such that the Klein effect can be
overcome.\cite{Rozhkov2011} Theoretically, a low energy effective
model that is well suited to describe various disorder phenomena has
been constructed by means of a chiral gauge
theory.\cite{Hou:2006qc,Jackiw2007} This has been supplemented by
scalar and gauge fields that account for the self-interaction of the
carbon background and the mean self-interaction of the Dirac
fermions.\cite{Oliveira:2010hq} In this framework, charge confinement
has been described by modeling one-dimensional defects, associated
e.g. with chemical bonding of foreign atoms to carbon atoms, as
potential barriers which break sublattice
symmetry.\cite{Popovici:2012xs}

The single-particle picture, however, changes quite dramatically if
the long range Coulomb interaction is taken into account. Since
fermions are confined to live in the plane, whereas the gauge bosons
live in three spatial dimensions, one is confronted with the problem
of coupling the two-dimensional current with a three-dimensional gauge
potential. This difficulty is solved by integrating the usual photon
propagator over the third momentum dimension (known as dimensional
reduction), which leads to an atypical behavior of the gauge
propagator, namely, it goes like $1/k$, instead of $1/k^2$. This
anomalous gauge propagator induces the renormalization of the electron
self-energy, which is not present in ordinary QED$_3$.

Various theoretical studies have predicted that Coulomb interaction
may generate dynamically a finite mass gap, such that graphene
undergoes a phase transition from semimetal to insulator above some
critical coupling
$\al_c$.\cite{Gamayun:2009em,khve2009,vafek2008,drula2009,Buividovich:2012nx}
This possibility is motivated by the large value of the 'bare'
coupling $\al=e^2/\epsilon h v_F$ of graphene on a substrate with
dielectric constant $\epsilon$ (for suspended graphene
$\alpha_0=2.19$).

On the other hand, most recent experiments find no signature of an
insulating phase in suspended graphene.\cite{elgo2012,mayorov2012}
Measurements indicate, however, that the real dispersion relation is
logarithmic, instead of linear, near the neutrality point. This
reshaping of the Dirac cones is caused by the charge-carrier density
dependent renormalization of the Fermi velocity (induced by the
Coulomb interaction). In fact, the running of the fermion velocity
with the energy was already predicted in earlier renormalization group
calculations,\cite{gonzalez1999} where it was found that the Fermi
velocity grows logarithmically without bound, until retardation
effects become important enough to invalidate the instantaneous
Coulomb approximation.

To analyze the problem of the gap generation, it is useful to inspect
the \DS equation for the fermion propagator \fig{fig:gapeq}. Following
the usual procedure, this equation is converted into a system of
coupled integral equations for the Fermi velocity dressing function
$A(p)$ and the gap function $\Delta(p)$, which characterize the
fermion propagator,
\be
S(p)^{-1}=\ga^0p_0-v_F A(p)\,\ga^ip_i-\Delta(p).
\ee

At the critical coupling $\al_c$, where the phase transition should
take place, the system is well described by bifurcation
theory. Applying bifurcation theory amounts to linearizing the
equations around the critical point, i.e., neglect quadratic and
higher order terms. By using a one-loop photon propagator, it has been
found that the function $A(p)$ is logarithmically
divergent,\cite{Popovici:2013fwp} in agreement with the experimental
evidences.\cite{elgo2012} Explicitly, it reads ($g=\al N_f\pi/4$)
\be
A(p)=1+\frac{2}{\pi^2N_f g}\left\{
-\left[\pi-2g+c(g)(g^2-1)\right]\ln\frac{p}{\La} +f(g) \right\}, 
\label{eq:Asolfull}
\ee
where $c(g)$ and $f(g)$ are functions of the coupling, $N_f=2$ for
monolayer graphene, and the physical cutoff $\Lambda $ is determined
by the size of the graphene's Brillouin zone. This result is then used
in the bifurcation analysis of the gap equation, which around the
critical point reduces to:
\be
\Delta(p)=\frac{e^2}{v_F \varepsilon}\int\frac{d^2\vec
  k}{(2\pi)^2}\frac{1}{k\, |\vec p- \vec k|}
\frac{\Delta(k)}{A(k)}
J\left (\frac{k}{ |\vec p- \vec k|} A(k),g\right). 
\label{eq:gapD2}
\ee
In the above, $J$ is a piecewise function that includes the effects of
the Fermi velocity dressing function. Importantly, for determining the
critical coupling corresponding to the semimetal-insulator transition,
not only the logarithmic but also the regular contributions to the
function $A(p)$, i.e. the function $f(g)$, become
significant.\cite{Popovici:2013fwp} For the critical coupling it is
obtained $\alpha_c=2.85$, which is larger than the bare coupling of
suspended graphene $\alpha_0=2.19$. These findings are in agreement
with the experimental observations that suspended graphene does not
undergo a phase transition.\cite{elgo2012} Hence, it appears that the
logarithmic renormalization of Fermi velocity strongly weakens the
Coulomb interaction, favoring the persistence of the semimetal phase
in suspended graphene.\cite{Popovici:2013fwp}

Despite the encouraging results, a complete understanding of the
physical picture requires a full nonperturbative investigation, by
including the \DS equation for the gauge propagator. Since
nonperturbative contributions to both photon propagator and vertex
functions have led to significant corrections in ordinary QED$_3$, it
is likely that including these effects will lead to further
corrections of the critical coupling. Furthermore, a proper comparison
of our value for the critical coupling with lattice calculations is
impeded by the details of the cutoff procedure in the ultraviolet, as
lattice simulations have been performed on both 'standard' squared
\cite{drula2009} and hexagonal (physical) \cite{Buividovich:2012uk}
lattices. Finally, as well known from ordinary QED$_3$, finite volume
effects may also play a role, see for example
Ref. \refcite{Goecke:2008zh} and references therein.

When comparing to the experiment, one has to keep in mind that other
type of terms, such as four-fermion contact interactions, may also
become important and thus have to be included in the model Lagrangian,
in addition to the long-range Coulomb interaction. First studies in
this direction have already been undertaken in
Refs.~\refcite{Gamayun:2009em,Mesterhazy:2012ei,Herbut:2009qb}. Another
important line of study is to explore the effects of a finite chemical
potential on the Fermi velocity --- experimentally this corresponds to
a finite density of charge carriers, i.e., chemical doping. This
problem is currently under investigation.\cite{inprep1}

%%%%%%%%%%%%%%%%%%%%%%%%%%%%%%%%%%%%%%%%%%%%%%%%%%%%%%%%%%%%%%
%%%%%%%%%%%%%%%%%%%%%%%%%%%%%%%%%%%%%%%%%%%%%%%%%%%%%%%%%%%%%%
\subsection {High-temperature superconductivity}
\label{subsec:HTS}

Due to their properties apparently incompatible with conventional
metal physics, high-T$_c$ cuprate superconductors can be regarded as
another example of QED$_3$ 'in the lab'. A particularly striking
feature is the pseudogap --- a third phase that lays at the boundary
between the nonsuperconducting (in this case, associated with the
antiferromagnetic spin density wave) and superconducting
phases.\cite{Herbut:2002yq} The physics of the pseudogap phase is
dominated by strong pairing fluctuations, as has been experimentally
demonstrated by measuring the so-called Nernst signal.\cite{Xu2000}

Describing the pseudogapped phase theoretically is a challenge due to
the complications introduced by the scattering of the fermions off the
vortices (topological defects), which are otherwise bounded in the
superconducting phase.\cite{Vafek2001} A way to circumvent this
problem has been put forward by Franz and Tesanovi{\'c}, who adopted
an 'inverted' paradigm and proposed to start from the 'conventional'
superconducting phase and try to find the mechanism that describes the
transition to the normal state.\cite{Franz:2001zz} In this approach,
the fermion field is gauge transformed to a new field which the
authors call 'topological fermions'. The pseudogap phase of the
topological fermions is then associated with the chiral symmetric
QED$_3$, whereas the transition to the antiferromagnetic phase is
indicated by the generation of a dynamical mass for the
quasiparticles.\cite{Franz:2002qy,Franz:2001zz,Herbut:2002yq,Herbut:2002wd}
The generation of such a mass term results from interactions of the
fermionic quasiparticles with topological excitations described by the
gauge fields of QED$_3$.

An open question is whether at zero temperature a pseudogap associated
to the onset of antiferromagnetic spin density wave instability is
formed in the first place, or if the system is driven directly into
the antiferromagnetic phase. The answer lays in the critical number of
'flavors' of fermions $N_f^c$ where the system undergoes a phase
transition from the chirally broken into the chirally symmetric
phase. This has to be compared with the physical number of flavors
$N_f=2$, which denotes the number of low energy quasiparticles located
at the four nodal points of the gap function, see for example
Ref.~\refcite{Franz:2002qy} for details. If $N_f^c<2$, the theory will
first go through the pseudogap phase; if, on the other hand,
$N_f^c>2$, the system is driven directly from the superconducting into
the antiferromagnetic phase upon further underdoping.

Furthermore, since experimental observations suggest that most of the
HTS materials are strongly anisotropic due to the large difference
between the Fermi velocity and a second velocity $v_{\Delta}$ (related
to the amplitude of the superconducting order
parameter),\cite{chiao2000,mesot1999} a complete theoretical
description requires taking into account this intrinsic anisotropy as
well. At the level of the Lagrangian, including the anisotropy amounts
to implementing a nontrivial metric that collects the anisotropic
fermion velocities.\cite{Bonnet:2011hh}

Much work has been devoted to determining the critical number of
flavors $N_f^c$ at which chiral symmetry is broken, in the isotropic
\cite{Appelquist:1988sr,Nash:1989xx,Maris:1996zg,fial2004} as well as
anisotropic
\cite{Franz:2002qy,Herbut:2002yq,Lee:2002qza,Bonnet:2011hh} versions
of QED$_3$. Studies carried out for isotropic QED$_3$ indicate that
$N_f^c$ lays around $3.5-4$. In the following, we concentrate on
anisotropic QED$_3$, and briefly review the results obtained within
the \DS formalism.  In a first study, the \DS equation for the fermion
propagator has been solved in the limit of small anisotropic
velocities.\cite{Franz:2002qy} It has been found that in this case
there should be essentially no difference between the universal
behavior of isotropic and anisotropic QED$_3$. Later on, a more
sophisticated fermion dressing has been included, and again only the
case of small anisotropic velocities has been
considered.\cite{Herbut:2002yq,Lee:2002qza} In the large-$N_f$ limit,
it has been found that the critical number of flavors where the theory
suffers the chiral phase transition stays the same. In a subsequent
study,\cite{Concha:2009zj} a quantitative criterion that relates the
critical number of fermion flavors and the strength of the
photon-fermion coupling has been proposed, namely, it was conjectured
that the dynamical fermion mass will be generated as soon the gauge
field strength is larger than some threshold value which can be
determined from the S-matrix for fermion-fermion scattering. It was
concluded that velocity anisotropy does affect the number of critical
fermion flavors at which chiral symmetry is broken due to mass
generation. A recent reexamination of the situation going beyond the
small anisotropy expansion, and using a power law ansatz for the
photon propagator, has shown sizable deviations of the critical number
of flavors from the isotropic case,\cite{Bonnet:2011hh,Bonnet:2012az}
in agreement with lattice results.\cite{Hands:2004ex,Thomas:2006bj}

%%%%%%%%%%%%%%%%%%%%%%%%%%%%%%%%%%%%%%%%%%%%%%%%%%%%
%%%%%%%%%%%%%%%%%%%%%%%%%%%%%%%%%%%%%%%%%%%%%%%%%%%%

\section{Summary}
\label{sec:concl}

In this brief review we have presented a few selected applications of
\DS approach to strongly interacting fermion systems. In the first
part we have considered the confinement problem, concentrating on the
heavy quark limit of Coulomb gauge QCD.  With the truncation to
include only dressed \YM propagators, we have sketched the derivation
of a simple relation between the Green's functions of Coulomb gauge
\YM theory and the quark confinement potential, for quark-antiquark
and three-quark systems. Confining (finite energy) solutions exist
only for color singlet meson/baryon states, and otherwise the systems
have infinite energy. Further, we have seen that in the heavy quark
limit, the rainbow approximation to the quark gap equation and the
ladder approximation to the associated \BS and Faddeev equations
remain nonperturbatively exact. In the remainder of this part we have
briefly reviewed Landau gauge results on the infrared \YM propagators.

In the second part we have moved on to less commonplace applications,
in graphene and high temperature superconductors. These planar
condensed matter systems are well described by effective QED$_3$-like
theories, which are tailored to accommodate the particularities of the
system under consideration. We have reviewed selected topics studied
with \DS methods, which include the problem of dynamical gap
generation by long-range Coulomb interactions in suspended graphene,
and the pseudogap phase in high temperature superconductors. In
graphene, including renormalization effects on the Fermi velocity has
lead to the conclusion that the Coulomb interaction is strongly
weakened near the charge neutrality point, preventing the emergence of
a gapped phase. These results agree with the experimental findings
which indicate the persistence of the semimetal phase. In
high-temperature superconductivity, where the physics of the pseudogap
is strongly dependent on the number of fermion species, it has been
shown that the inherent spatial anisotropy leads to a significant
alteration of the critical number of flavors where the phase
transition from chirally symmetric to chirally broken phase takes
place (as compared to isotropic QED$_3$). Least but not last, it is
important to appreciate the possible similarities between different
theories, i.e., the scaling solutions of Landau gauge propagators of
\YM theory and QED$_3$. In the context of high-T$_c$
superconductivity, a power law ansatz for the photon propagator has
been successfully employed to study the dynamical breaking of chiral
symmetry.  Whether the gauge field in graphene is characterized by a
similar behavior remains to be clarified in the future.

\section*{Acknowledgments}

The author is grateful to Christian~Fischer and Stefan~Strauss for
critical readings and many helpful comments on the manuscript. This
work was supported by the Deutsche Forschungsgemeinschaft within the
SFB 634 and the Helmholtz International Center for FAIR within the
LOEWE program of the State of Hesse.

%\ifgeneratebbl
%\bibliographystyle{utphys-min}
%\bibliography{$HOME/bibliography/biblio}
%\else
%\input{mpla-rev.bbl}
%\fi

%\end
\providecommand{\href}[2]{#2}

%\begingroup\raggedright

%\endgroup

\end{document}